\documentclass[preprint,12pt]{elsarticle}

\usepackage{url}
\usepackage{amsmath}
\usepackage{amssymb}
\usepackage{graphicx}
\usepackage{float}
\usepackage [english]{babel}
\usepackage [autostyle, english = american]{csquotes}
\MakeOuterQuote{"}
\usepackage{dcolumn}
\usepackage{comment}
\usepackage[usenames,dvipsnames]{color}
\usepackage{caption}
\usepackage{bm}

\journal{Chaos, Solitons $\&$ Fractals}

\begin{document}

\begin{frontmatter}



\title{Stabilization of self-steepening optical solitons in  a periodic $\mathcal{P}\mathcal{T}$-symmetric  potential}


\author[inst1]{Eril Güray Çelik\corref{cor1}}%
\ead{celik19@itu.edu.tr}
\cortext[cor1]{Corresponding author}

\author[inst1]{Nalan Antar}

\affiliation[inst1]{organization={Department of Mathematics Engineering, Istanbul Technical University, Maslak},
            city={Istanbul},
            postcode={34469}, 
            country={Türkiye}}

\begin{abstract}
We numerically investigate the existence and stability dynamics of self-steepening optical solitons in a periodic $\mathcal{P}\mathcal{T}$-symmetric potential.
 We show that self-steepening solitons of the modified nonlinear Schrödinger (MNLS) equation undergo a position shift and amplitude increase during their evolution in the MNLS equation. The stabilization of solitons by an external potential is a challenging issue. This study demonstrates that the suppression of both the amplitude increase and the position shift of self-steepening solitons can be achieved by adding a periodic $\mathcal{P}\mathcal{T}$-symmetric potential to the MNLS equation.
\end{abstract}


\end{frontmatter}


\section{\label{sec:Int}Introduction}

Optical solitons are solitary waves that arise from a delicate balance between the group velocity dispersion (GVD)  and nonlinear effects caused by the optical Kerr effect \cite{hasegawa_tappert_1973_1,mahalingam_porsezian_2001}. The generation and analysis of solitons in optics is a fairly popular research topic since optical solitons have a wide range of applications, including femtosecond lasers \cite{salin_grangier_roger_brun_1986}, logic gates and filters in optical logic devices \cite{Swartzlander:92,Geng_2022}, pulse compression and splitting in ultrafast optics \cite{mollenauer_stolen_gordon_1980}, and all-optical switching \cite{vicencio_molina_kivshar_2004}. In particular, the propagation of optical solitons in fiber-optic communication systems is an area of great interest for research because of their remarkable stability properties \cite{krökel_halas_giuliani_grischkowsky_1988, Xu_2023, Wen_2023}.

 Optical solitons can propagate through long distances in fiber transmission systems without being affected by chromatic and polarization mode dispersion. Since their natural structure is preserved, they can be used as natural optical bits of information in fiber optic systems. 
 In mono-mode optical fibers, the envelope of an electromagnetic field of a light signal contains the signal information and changes slowly because it has a high-frequency carrier wave. Also, since the refractive index to which the light signal is exposed in the fiber depends on the intensity of the light, this phenomenon is called the Kerr nonlinear effect, the light signal has a weak nonlinearity.   Hence, light pulse propagation in mono-mode optical fibers is governed by the nonlinear Schrödinger (NLS) equation \cite{hasegawa_tappert_1973_1,yangbook}.  
 However, the classical NLS  equation may be inadequate to model the physical system for femtosecond pulses. This is because when the width of the light pulse is very small, i.e., the frequency is high,  the higher-order NLS equation has to be taken into account, which includes some higher-order effects such as third-order dispersion,  self-steepening, and stimulated  Raman scattering   \cite{kodama_hasegawa_1987, singer_potasek_fang_teich_1992, AGRAWAL2019127,horikis2019perturbations,Liu_2024, Li_2024}. 
 The impact of these higher-order effects on soliton propagation in nonlinear optical fibers has been the subject of some studies before \cite{mahalingam_porsezian_2001, sasa_narimasa_1991,nakkeeran_porsezian_sundaram_mahalingam_1998,DONG19968,GROMOV2020105220,Zhu_2022}. The self-steepening is one of the most important of these higher-order effects and besides soliton propagation in fiber optics, it has significant applications such as the propagation of nonlinear Alfvén waves in plasmas \cite{mio265,mjølhus_1976} and light-matter interactions \cite{moses2007self}. 
 Self-steepening becomes crucial for the propagation of short pulses in long optical fibers or waveguides    \cite{tzoar_jain_1981, anderson1983nonlinear}. The propagation of an optical soliton under the self-steepening effect can be described by the modified NLS (MNLS) equation 
 \begin{equation}\label{mnlse}
	i u_{z}+\frac{\beta}{2} u_{x x} +\left|u\right|^{2} u + is \frac{\partial}{\partial x}\left(|u|^{2}u\right)=0, 
 \end{equation}
{where complex-valued function $u(x,z)$ is the envelope of the light's electric field, $x$ is the transverse coordinate, $z$ is the distance along the direction of propagation,  $\beta$ is the GVD coefficient, and $s$ is the self-steepening coefficient.} 
The self-steepening effect occurs when the group velocity of an optical pulse depends on the density \cite{demartini_townes_gustafson_kelley_1967,han_park_2011}. In such a case, since the group velocity is density-dependent, the peak of an optical pulse moves slower than its wings \cite{mahalingam_porsezian_2001}. Therefore, the self-steepening causes the top of an optical pulse to become steeper towards the trailing edge and causes an optical pulse to become asymmetric \cite{anderson1983nonlinear, Anderson:82, Yin:07}. In other words, it causes an optical shock at the trailing edge, especially in the absence of the GVD effect \cite{anderson1983nonlinear, han_park_2011, Anderson:82, deOliveira:92}. The GVD  dampens the effects of the self-steepening to a remarkable amount \cite{anderson1983nonlinear,han_park_2011}. However, despite the GVD, the self-steepening severely affects the stability of the soliton by causing a shift in the position of the pulse and dividing the pulse into sub-pulses \cite{deOliveira:92}. These impacts of the self-steepening on an NLS soliton are well-known phenomena.
{ In this paper, we examine the dynamic properties of self-steepening solitons of the MNLS equation and show that these self-steepening solitons undergo a position shift and amplitude increase during their evolution in the MNLS equation, leading to inherent instability. This finding highlights a critical challenge for practical applications, as unstable solitons can distort information and limit transmission distances in fiber-optic communication systems. We present a novel solution to address this challenge: incorporating a periodic $\mathcal{P}\mathcal{T}$-symmetric potential into the MNLS equation.  We demonstrate that the suppression of both the amplitude increase and the position shift of self-steepening solitons can be achieved by adding a periodic $\mathcal{P}\mathcal{T}$-symmetric potential to the MNLS equation.}

${\mathcal{P} \mathcal{T}}$-symmetric quantum systems were discovered in 1998 by Bender and Boettcher, who proposed the idea that Hermitian Hamiltonians could be extended to non-Hermitian Hamiltonians \cite{bender_boettcher_1998}. After that, $\mathcal{P}\mathcal{T}$-symmetric systems became the subject of research in many fields such as microwave cavities \cite{dietz2011exceptional}, electronic circuits \cite{schindler2011experimental}, lasers \cite{chong2011p}, chaos and noise \cite{schomerus2010quantum} and optics \cite{ruter_makris_2010,regensburger2012parity}. In 2008, Musslimani et al. studied the existence, stability, and propagation dynamics of solitons in $\mathcal{P}\mathcal{T}$-symmetric lattices \cite{musslimani_makris_el-ganainy_christodoulides_2008}. Subsequently, many studies have been conducted on $\mathcal{P}\mathcal{T}$ solitons as certain characteristics of solitons in optical lattices can be controlled by adjusting the lattice depth and period \cite{makris_el-ganainy_christodoulides_musslimani_2011,zeng_lan_2012,burlak_malomed_2013,ge_shen_zang_ma_dai_2015,goksel_antar_bakirtas_2015}. In this study, we use $\mathcal{P}\mathcal{T}$-symmetric periodic lattices to stabilize self-steepening solitons.


The structure of the paper is as follows: Section 2 presents the governing equation (model) describing the propagation of self-steepening solitons in a periodic ${\mathcal{P} \mathcal{T}}$-symmetric potential. The governing equation cannot be solved analytically.  Section 3 introduces the pseudospectral renormalization (PSR) method employed to obtain self-steepening solitons numerically.  The existence region of self-steepening solitons is investigated in Section 4. The linear and nonlinear stability analysis of self-steepening solitons and numerical results are depicted in Sections 5 and 6. A summary of the numerical results is given in Section 7.

\section {Theoretical Model}

In fiber optic communication systems, the self-steepening effect significantly impacts optical solitons when pulse durations are shortened to the femtosecond regime to achieve high bandwidth. The propagation of self-steepening optical solitons in a periodic ${\mathcal{P} \mathcal{T}}$-symmetric potential is governed by the following equation
\begin{equation}\label{ps1}
	i u_{z}+\frac{\beta}{2} u_{x x} +\left|u\right|^{2} u +is \frac{\partial}{\partial x}\left(|u|^{2}u\right)+V_{\mathcal{P} \mathcal{T}}(x) u=0, 
 \end{equation}
where $\frac{\partial}{\partial x}\left(|u|^{2}u\right)$ is the self-steepening term, and $V_{\mathcal{P} \mathcal{T}}(x)=V(x)+iW(x)$ represents a periodic ${\mathcal{P} \mathcal{T}}$-symmetric potential. The real and imaginary parts of a ${\mathcal{P} \mathcal{T}}$-symmetric potential satisfy the conditions $V(-x)=V(x)$ and $W(-x)=-W(x)$.  Physically, $V(x)$ corresponds to the spatial distribution of the refractive index, and the $W(x)$ corresponds to the balanced gain-loss relationship. We consider the following  periodic ${\mathcal{P} \mathcal{T}}$-symmetric potential 
\begin{equation}\label{pt}
	V_{\mathcal{P} \mathcal{T}}(x)=V_{0} \cos ^{2}(x)+i W_{0} \sin (2 x).
\end{equation}
Here, $V_0$ and $W_0$ are the depths of the real and imaginary parts of the potential, respectively. 


 Since Eq. (2) (with the potential given by Eq. (3)) cannot be solved analytically, we employ the PSR method \cite{antar_2014}, a robust numerical technique for such nonlinear equations, to identify localized solutions. 


\section{Pseudospectral Renormalization Method}
The spectral renormalization method  \cite{ablowitz_musslimani_2005}, originally derived from the Petviashvili method \cite{petviashvili_1976,ALVAREZ20142272}, is a powerful tool for numerically computing solitons in nonlinear waveguides.
This method is based on transforming the governing equation into Fourier space and determining a convergence factor with a nonlinear nonlocal integral equation coupled to an algebraic equation. The spectral renormalization method is easy to implement and usually converges fairly quickly. On the other hand, if the system lacks any homogeneity, as in the case of saturable nonlinearity, the convergence factor cannot be found explicitly by the spectral renormalization method. To determine the convergence factor, a root-finding method should be employed. In \cite{antar_2014}, the
pseudospectral renormalization (PSR) method is derived from the spectral renormalization method. This derivation allows for the explicit calculation of the convergence factor, even in the absence of homogeneity \cite{Göksel201883}.

 We utilize a modification of the PSR method to find the localized solutions of Eq. (2). We seek a soliton solution of Eq. (\ref{ps1}) in the form 
 \begin{equation} \label{soliton}
 u(x, z)=f(x) e^{i \mu z},
\end{equation}
 where $f(x)$ is a localized complex-valued function, and $\mu>0$ is the propagation constant. 
Substituting this solution ansatz into   Eq. (\ref{ps1}), we get the following nonlinear eigenequation for $f$ and $\mu$
\begin{equation}\label{ps5}
		-\mu f +\frac{\beta}{2} \frac{{d^2 f}}{{dx^2 }}+|f|^{2} f + V_{\mathcal{P} \mathcal{T}} f + 
		i s \frac{d}{d x}\left( f |f|^{2}\right) =0,
\end{equation}
with the boundary condition $f \rightarrow 0$ as $|x| \rightarrow+\infty$.
Applying the Fourier transform and inverse Fourier transform  to Eq. (\ref{ps5}), respectively, gives the following equation 
\begin{equation}\label{ps6}
 - \mu f - \frac{\beta }{2}{\cal F}^{ - 1} \left\{ {k^2 \hat f} \right\} + |f|^2 f + V_{{\cal P}{\cal T}} f + is{\cal F}^{ - 1} \left\{ {ik{\cal F}\left\{ {f|f|^2 } \right\}} \right\} = 0.
\end{equation}
Here $\mathcal{F} $ and $  \mathcal{F}^{-1}$  denote the Fourier and inverse Fourier transforms, respectively, and  $k$ is the Fourier variable.  $\mu f$ can be written as
\begin{equation}\label{ps66}
    \mu f=\mathcal{F}^{-1} \left\{\mu \hat{f}\right\}.
\end{equation}
Substituting Eq. (\ref{ps66}) into Eq. (\ref{ps6}), we get
\begin{equation}\label{ps7}
\mathcal{F}^{-1}\left\{(\mu +\frac{\beta}{2} k^{2}) \hat{f} \right\}
  =|f|^{2} f+
V_{\mathcal{P} \mathcal{T}} f+
i s  \mathcal{F}^{-1}\left\{i k \mathcal{F}\left\{f |f|^{2}\right\}\right\}.
\end{equation}
If we employ the fixed-point iteration method to find a localized solution of 
Eq. (\ref{ps7}), the solution either grows without bound or tends to zero under iteration. To avoid this situation, we should introduce a new field variable, $f(x)=\lambda w(x)$, where $\lambda$ is a real-valued constant to be determined and called the convergence factor. Substituting this new variable into Eq. (\ref{ps7}), the function $w(x)$ satisfies
\begin{equation}\label{ps8}
{\cal F}^{ - 1} \left\{ {(\mu  + \frac{\beta }{2}k^2 )\hat w} \right\} = {\rm{ }}|\lambda |^2 |w|^2 w + V_{{\cal P}{\cal T}} w + is|\lambda |^2 {\cal F}^{ - 1} \left\{ {ik{\cal F}\left\{ {w|w|^2 } \right\}} \right\}.
\end{equation}
Multiplying Eq. (\ref{ps8})  by  ${w}^{*}$ and integrating over ($-\infty,\infty$),
we obtain an algebraic equation for the convergence factor
\begin{equation}\label{ps9}
\left| \lambda  \right|^2  = \frac{{S_1 }}{{S_2 }},
\end{equation}
where $S_1$ and $S_2$ are defined  by
\begin{equation}
\begin{array}{l}
 S_1  = \int_{ - \infty }^\infty  {\left[ {\mu ww^*  + \frac{\beta }{2}w^* {\cal F}^{ - 1} \left\{ {k^2 \hat w} \right\} - V_{{\cal P}{\cal T}} (x)ww^* } \right]} dx, \\ \\
 S_2  = \int_{ - \infty }^\infty  {w^* } \left[ {|w|^2 w + is{\cal F}^{ - 1} \left\{ {ik{\cal F}\left\{ {w|w|^2 } \right\}} \right\}} \right]dx. \\ 
 \end{array}
\end{equation}
The desired solution can be obtained by iterating Eq. (\ref{ps8}) in Fourier space
\begin{equation}\label{iter}
\hat w_{n + 1}  = \frac{{F\left\{ {|\lambda _n |^2 |w_n |^2 w_n  + V_{{\cal P}{\cal T}} w_n  + isw_n {\cal F}^{ - 1} \left\{ {ik{\cal F}\left\{ {|\lambda _n |^2 w_n |w_n |^2 } \right\}} \right\}} \right\}}}{{\mu  + \frac{{\beta k^2 }}{2}}}.
\end{equation}
We start implementing Eq. (\ref{iter})  by choosing an initial function as $w_{0}=e^{-x^{2}}$, which yields $\lambda_0$ from Eq. (\ref{ps9}). Then, from Eqs. (\ref{iter}) and (\ref{ps9}), we obtain $\hat w_1$ and $\lambda_1$, respectively.
The iteration continues until two conditions are met
\begin{equation}   
\begin{array}{l}
 e_n^{(1)}  = \left\| {f_n  - f_{n - 1} } \right\|_{L_\infty  }\leq10^{-10},\\ \\ 
 e_n^{(2)}  = \left\| \begin{array}{l}
  - \mu f_n  - \frac{\beta }{2}F^{-1} \{ k^2 \hat f_n \} +  |f_n |^2 f_n  + V_{{\cal P}{\cal T}} f_n \\
  + is F^{ - 1} \left\{ ikF\left\{ {f_n |f_n |^2 } \right\} \right\} 
 \end{array} \right\|_{L_\infty}\leq10^{-10},
 \end{array}
\end{equation}
where $f_n=\lambda_n w_n$.

Figs. 1(a) and  1(b) show the real and imaginary parts of a self-steepening soliton solution of Eq. (2) (computed by the PSR method), respectively, while Fig. 1(c) displays the error diagrams ($e_n^{(1)}$ and $e_n^{(2)}$) of the PSR method.  We see that errors defined by $e_n^{(1)}$ and $e_n^{(2)}$  drop below $10^{-10}$ after $37$ PSR iterations.
\begin{figure}[H]
\includegraphics[width=\textwidth,keepaspectratio=true]{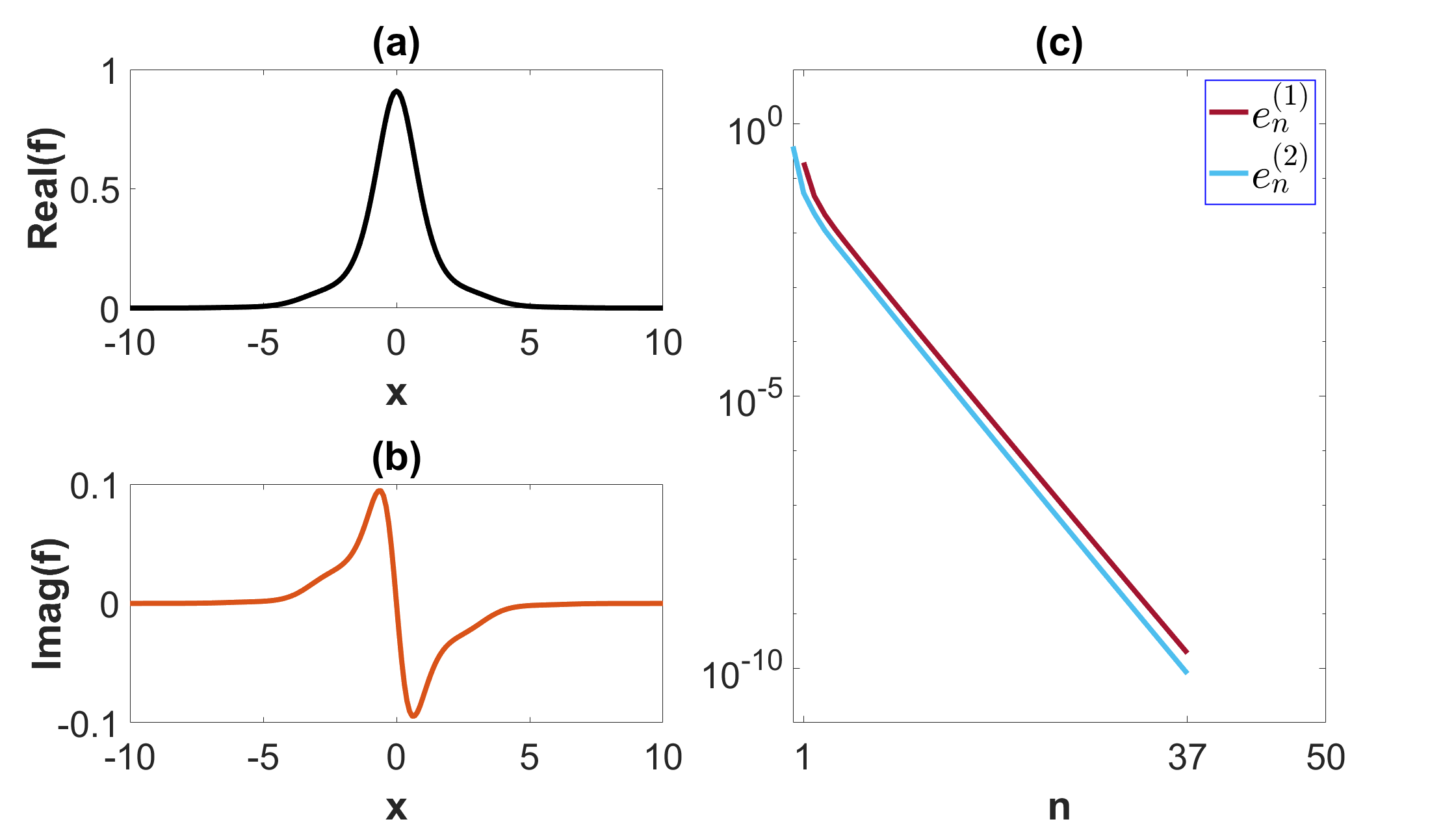}
\centering
\caption{Computation of a self-steepening soliton in Eq. (2)  (with $s=0.3, \beta=1, V_0=1, W_0=0.1, \mu=1$) by the PSR method. (a) and (b) Profiles of the real and imaginary parts of the soliton. (c) Graphs of $e_n^{(1)}$ and $e_n^{(2)}$  versus the number of iterations. }
\label{soliton_s03_v1_w01}
\end{figure}
\section{Existence of self-steepening optical solitons}
 
 We numerically investigate the existence region of self-steepening solitons of Eq. (2) according to the equation parameters.  Firstly, we study the effect of  GVD and 
 periodic ${\mathcal{P} \mathcal{T}}$-symmetric potential (3) on the existence of self-steepening solitons.
Fig. 2(a)  shows the existence region (shown in green) of soliton solutions of Eq. (2) (with $V_{\mathcal{P} \mathcal{T}}=0$) according to the self-steepening coefficient ($s$) and GVD coefficient ($\beta$) for four different values of $\beta$. As can be seen from the figure, the GVD dampens the self-steepening effect. Namely, as the $\beta$ coefficient increases, soliton solutions of Eq. (2) can be obtained for larger values of $s$. To scrutinize the effect of the periodic  ${\mathcal{P} \mathcal{T}}$-symmetric potential on the existence of self-steepening solitons, the relationship between $s$ and $\beta$ is reconsidered by taking the coefficients of the real and imaginary parts of the potential as $0.7$ and $0.1$, respectively, in Fig. 2(b).  Our analysis of the figure demonstrates that the periodic ${\mathcal{P} \mathcal{T}}$-symmetric potential significantly extends the existence region of self-steepening solitons. This implies that self-steepening solitons can be found for substantially larger values of the self-steepening coefficient compared to the case without the potential.
\begin{figure}[H]
\includegraphics[width=0.8\textwidth,keepaspectratio=true]{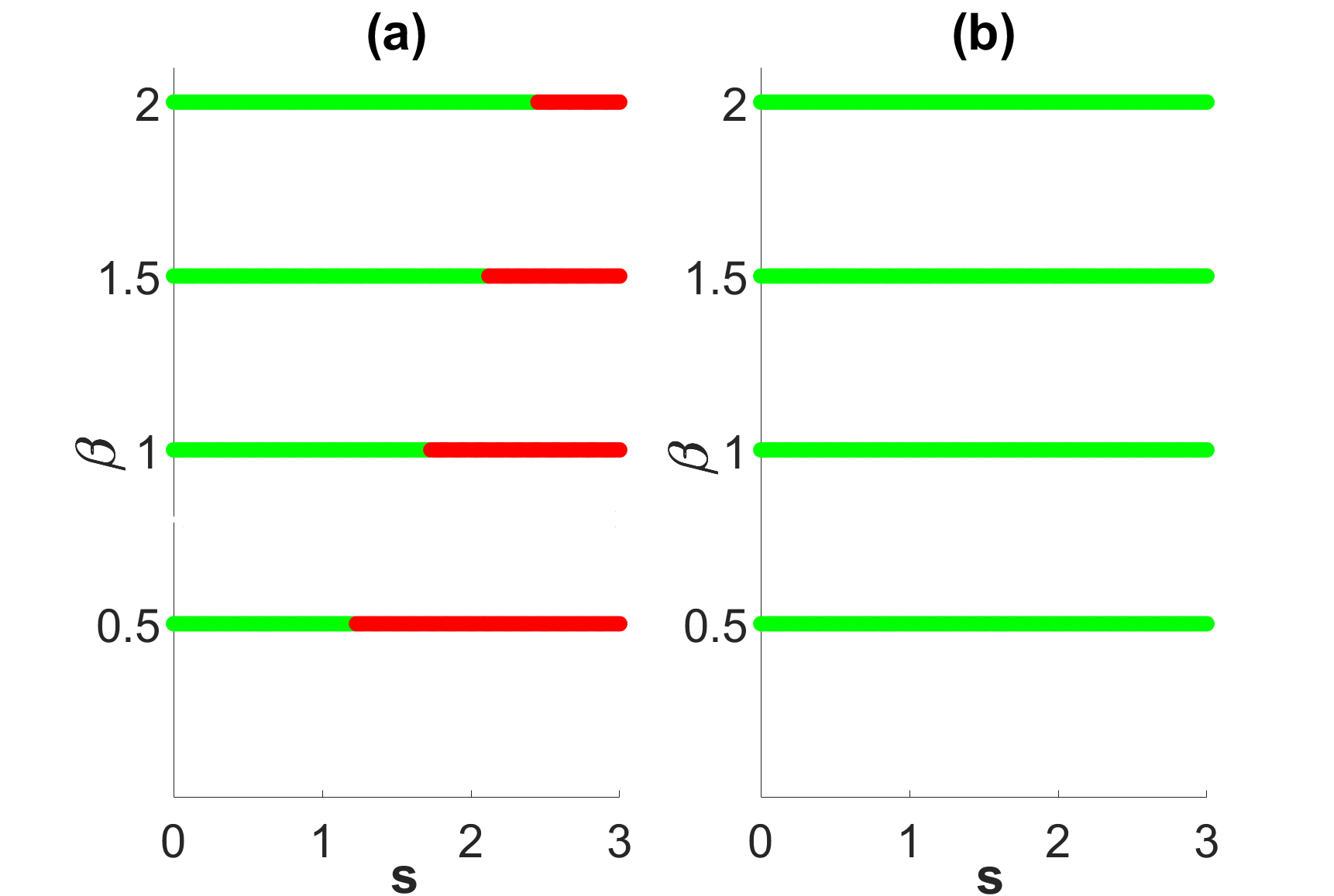}
\centering
\caption{The domain of existence (shown in green) for soliton solutions (with $\mu=1$) of Eq. (2) according to $s$ and $\beta$. (a)  $V_0=0$, $W_0=0$; (b)  $V_0=0.7$, $W_0=0.1$.}
\label{ch5_beta_s}
\end{figure}
%
%
Secondly, we investigate how the relationship between the coefficient of the real part of the periodic ${\mathcal{P} \mathcal{T}}$-symmetric potential and the propagation constant ($\mu$) affects the existence of self-steepening solitons.
In Fig. \ref{V0_mu_s_0_01_03_i_sp_2}, Eq. (2) (with $\beta=1$, $s=W_0=0.1$) has a soliton
solution for the values of $V_0$ and $\mu$ corresponding to the green region, while it has no soliton solution for the values corresponding to the red region. This figure reveals that increasing the value of $\mu$ enables the existence of self-steepening solitons for a wider range of $V_0$ values. 

To ensure consistent analysis of the existence and stability of self-steepening solitons, we fix the propagation constant ($\mu$) at $1$ throughout the paper. Consequently, our investigation focuses on the $V_0$ coefficient within the range of $0$ to $1$. This choice is motivated by the observation that larger values of $V_0$ exceeding $1$ necessitate a corresponding increase in $\mu$ to maintain soliton solutions.
\begin{figure}[H]
\includegraphics[width=0.8\textwidth,keepaspectratio=true]{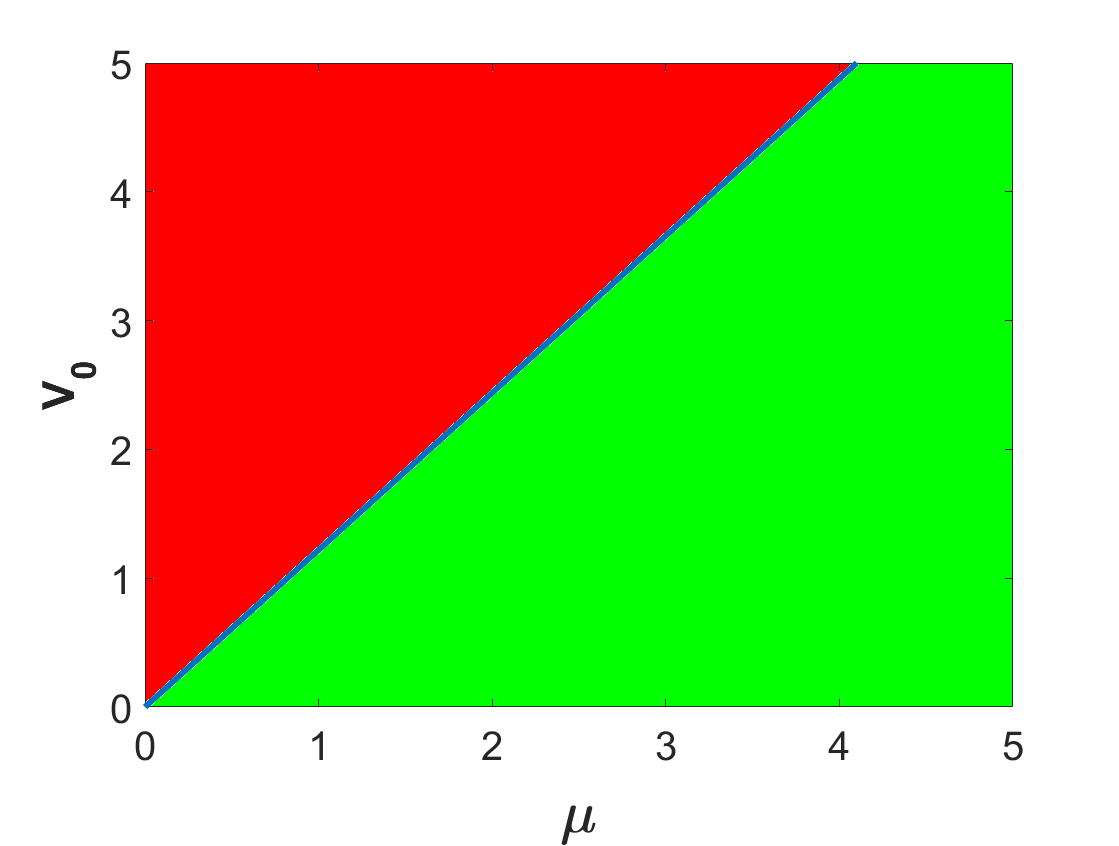}
\centering
\caption{The domain of existence (shown in green) for soliton solutions of Eq. (2) (with $s=W_0=0.1$)  according to $\mu$ and $V_0$.  }
\label{V0_mu_s_0_01_03_i_sp_2}
\end{figure}

 \section{Linear Stability of Self-Steepening Optical Solitons}

 In this section, we investigate the linear stability of self-steepening solitons in Eq. (2) by analyzing their linear stability spectra. To obtain a linear stability eigenvalue problem, we consider a perturbed solution of the form
  \begin{equation}
 \tilde{u}(x, z)=\left[f(x)+ \epsilon (g(x) e^{\lambda z}+h^{*}(x) e^{\lambda^{*} z })  \right] e^{i \mu z},
 \end{equation}
 where $g(x)$ and $h(x)$ are  perturbation eigenfunctions, $\lambda$ is the eigenvalue, and the superscript  "$^*$" represents complex conjugation. Substituting the perturbed solution into Eq. (\ref{ps1}) and linearizing,  we obtain the following linear stability eigenvalue problem for the self-steepening soliton 
 $u(x, z)=f(x) e^{i \mu z}$

\begin{equation}\label{eigv}
i\left[\begin{array}{cc}
\frac{1}{2} \frac{d^2}{d x^2}+G_0 \frac{d}{d x}+G_1 & G_2 \frac{d}{d x}+G_3 \\\\
-\left(G_2^* \frac{d}{d x}+G_3^*\right) & -\left(\frac{1}{2} \frac{d^2}{d x^2}+G_0^* \frac{d}{d x}+G_1^*\right)
\end{array}\right]\left[\begin{array}{l}
g \\\\
h
\end{array}\right]=\lambda\left[\begin{array}{l}
g \\\\
h
\end{array}\right],
\end{equation}
where
\begin{equation}
\begin{array}{ll}
G_0=2 i s|f|^2, & G_1=-\mu+V_{\mathcal{P} \mathcal{T}}+2|f|^2+4i s|f||f|_x , \\\\
G_2=i s f^2, & G_3=f^2+2 i s f f_x .
\end{array}
\end{equation}
We compute the whole stability spectrum of the soliton by solving  Eq. (\ref{eigv}) with the aid of the Fourier collocation method \cite{boyd_2001_chebyshev}.\footnote{A detailed explanation of the Fourier collocation method can be found in \ref{appendix}.}
 The linear stability spectrum ($\lambda$ spectrum) gives essential information about the behavior of the soliton under small perturbations. If the real part of any eigenvalues found in the linear spectrum is positive, the soliton is linearly unstable. Generally, as the positive real part of an eigenvalue increases, the perturbation growth rate of the soliton also increases. On the other hand, if the eigenvalues in the spectrum are purely imaginary, then some oscillations occur. In such a scenario, the soliton can be regarded as linearly stable.

\begin{figure}[H]
\includegraphics[width=\textwidth,keepaspectratio=false]{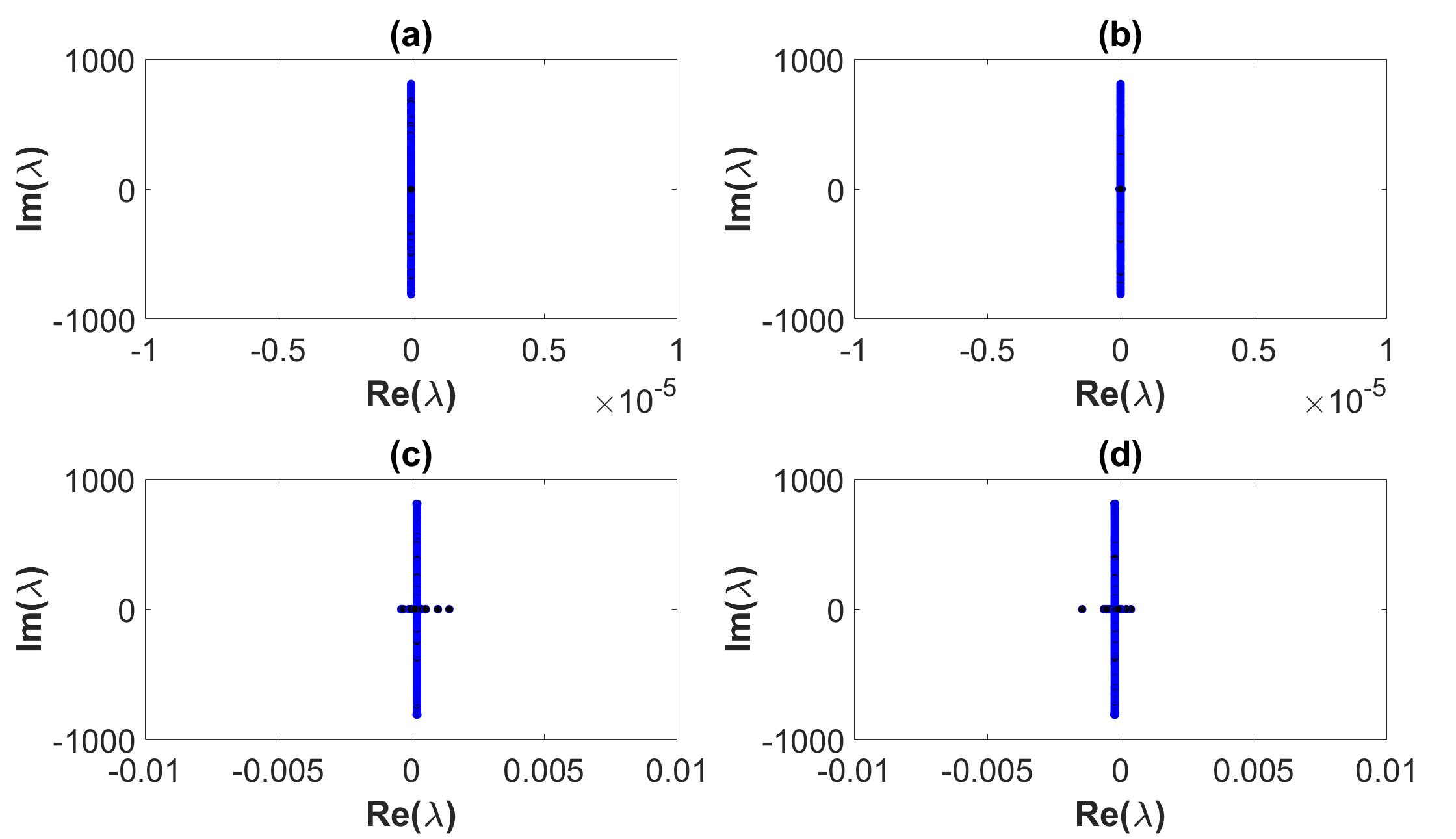}
\centering 
\caption{Linear stability spectra of self-steepening solitons of Eq. (2) (with $s=0.1$, $\beta=1$, $\mu=1$) in Eq. (2). (a)  $V_0=0, W_0=0$; (b)  $V_0=0.7, W_0=0$; (c)  $V_0=0.7, W_0=0.3$; (d)  $V_0=0.7, W_0=-0.3$.}
\label{linstb2_s01}
\end{figure}

\begin{figure}[H]
\includegraphics[width=\textwidth,keepaspectratio=false]{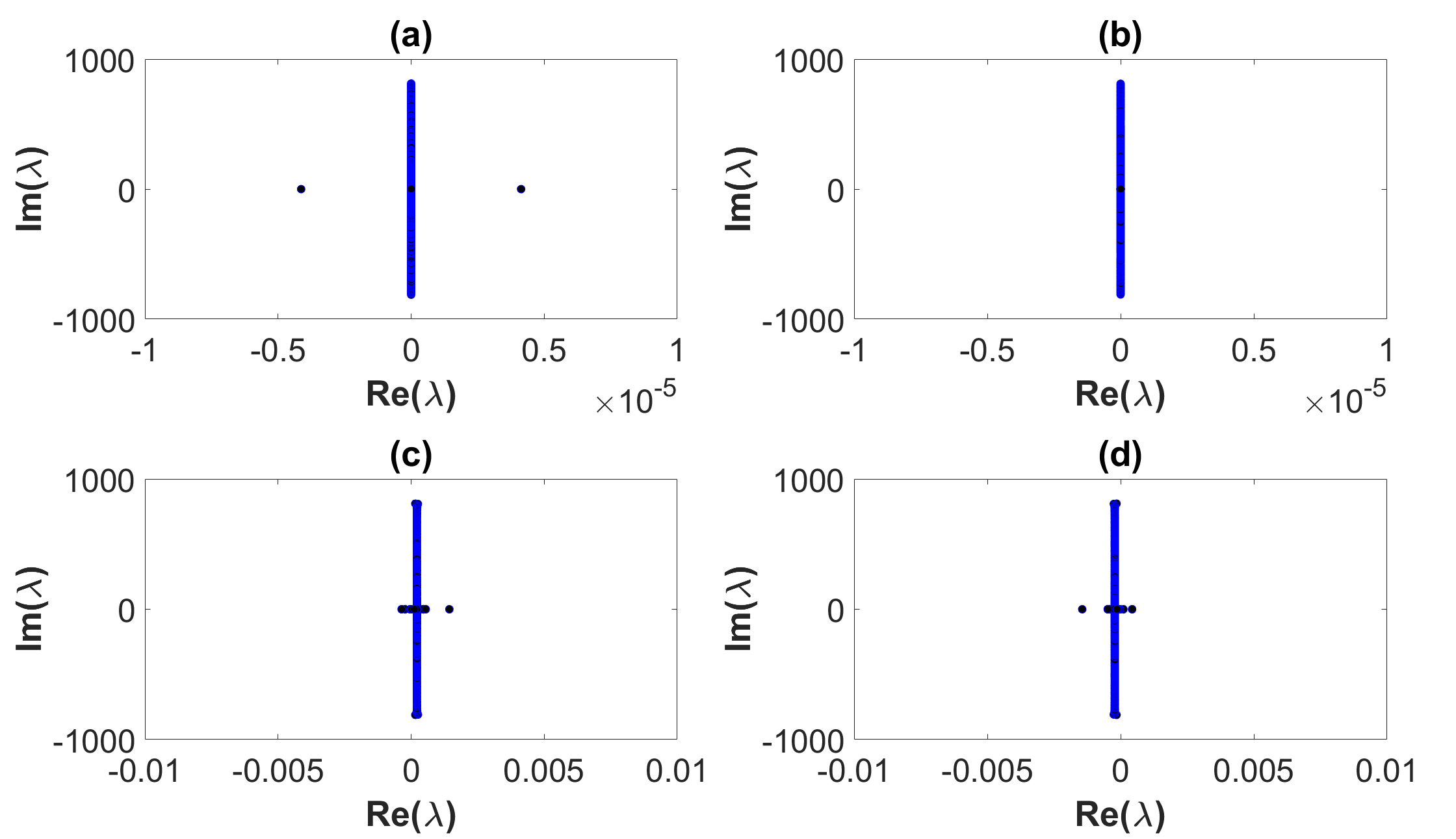}
\centering 
\caption{Linear stability spectra of self-steepening solitons of Eq. (2) (with $s=0.3$, $\beta=1$, $\mu=1$) in Eq. (2). (a)  $V_0=0, W_0=0$; (b)  $V_0=0.7, W_0=0$; (c)  $V_0=0.7, W_0=0.3$; (d)  $V_0=0.7, W_0=-0.3$.}
\label{linstb_s03}
\end{figure}

 Taking $s=0.1$ and $\beta=\mu=1$,  the spectra of self-steepening solitons for different depths of the real and imaginary parts of the periodic ${\mathcal{P} \mathcal{T}}$-symmetric potential (\ref{pt}) are displayed in Fig. \ref{linstb2_s01}. Fig. \ref{linstb2_s01}(a) shows the spectrum in the potential-free state ($V_0=W_0=0$). Since the spectrum lacks any eigenvalues with a positive real part, it can be inferred that the self-steepening soliton is linearly stable.  Additionally, we have observed that the spectrum contains no discrete eigenvalues other than a zero eigenvalue of multiplicity four. The absence of nonzero discrete eigenvalues in the linearization spectrum of solitons is a common characteristic of integrable equations \cite{mihalache_truta_panoiu_1993,kaup1978exact}.
 Next, it's shown that when the real part of the potential is added to the system ($V_0=0.7, W_0=0$), the resulting soliton is linearly stable, too; see Fig. \ref{linstb2_s01}(b). However, in this instance, it is worth noting that the spectrum includes certain internal modes. This presence is indicative of nonintegrable equations.  Subsequently, Fig.  \ref{linstb2_s01}(c) and  Fig. \ref{linstb2_s01}(d) illustrates the spectra of self-steepening solitons in the periodic  ${\mathcal{P} \mathcal{T}}$-symmetric potential. In these figures, the system has gain ($V_0=0.7, W_0=0.3$) and loss profiles ($V_0=0.7, W_0=-0.3$), respectively.  In both cases, self-steepening solitons are weakly
unstable, where the maximal linear instability growth rates of
solitons are very small ($\sim10^{-3}$ and $\sim3\times10^{-4}$, respectively). 
When the coefficient of self-steepening is increased to $0.3$ in the absence of potential, a pair of eigenvalues bifurcate out from the origin along the real axis, and the spectrum contains a positive real eigenvalue of multiplicity two; see Fig. \ref{linstb_s03}(a). However, since this positive real eigenvalue is very small ($\sim10^{-6}$), the soliton is linearly stable. Furthermore, the linear spectra of self-steepening solitons (with $s=0.3$) obtained with the addition of the periodic  ${\mathcal{P} \mathcal{T}}$-symmetric potential are similar to those observed for the case of $s = 0.1$. (Figs. \ref{linstb_s03}(b)-(d)).

\section{Nonlinear Stability of Self-Steepening Optical Solitons}
The nonlinear stability of self-steepening solitons is investigated through evolution simulations of Eqs. (1) and (\ref{ps1}) for long distances. To carry out these numerical investigations, we use the split-step method \cite{boyd_2001_chebyshev, strang_1968,yanenko_1971,yoshida_1990}. We show how the self-steepening term and the varying depths of the real and imaginary parts of the periodic $\mathcal{P}\mathcal{T}$-symmetric potential (\ref{pt}) affect the nonlinear stability of solitons. During the nonlinear stability analysis, we investigate whether the shape, position, and peak amplitude of self-steepening solitons change as they propagate in the $z$ direction. If the shape, position, and peak amplitude of the solitons do not change during the simulation, or if the changes are extremely small, they are considered to be nonlinearly stable.

\subsection{Self-steepening optical solitons without an external potential } 

\begin{figure}[ht]
\includegraphics[width=\textwidth,keepaspectratio=true]{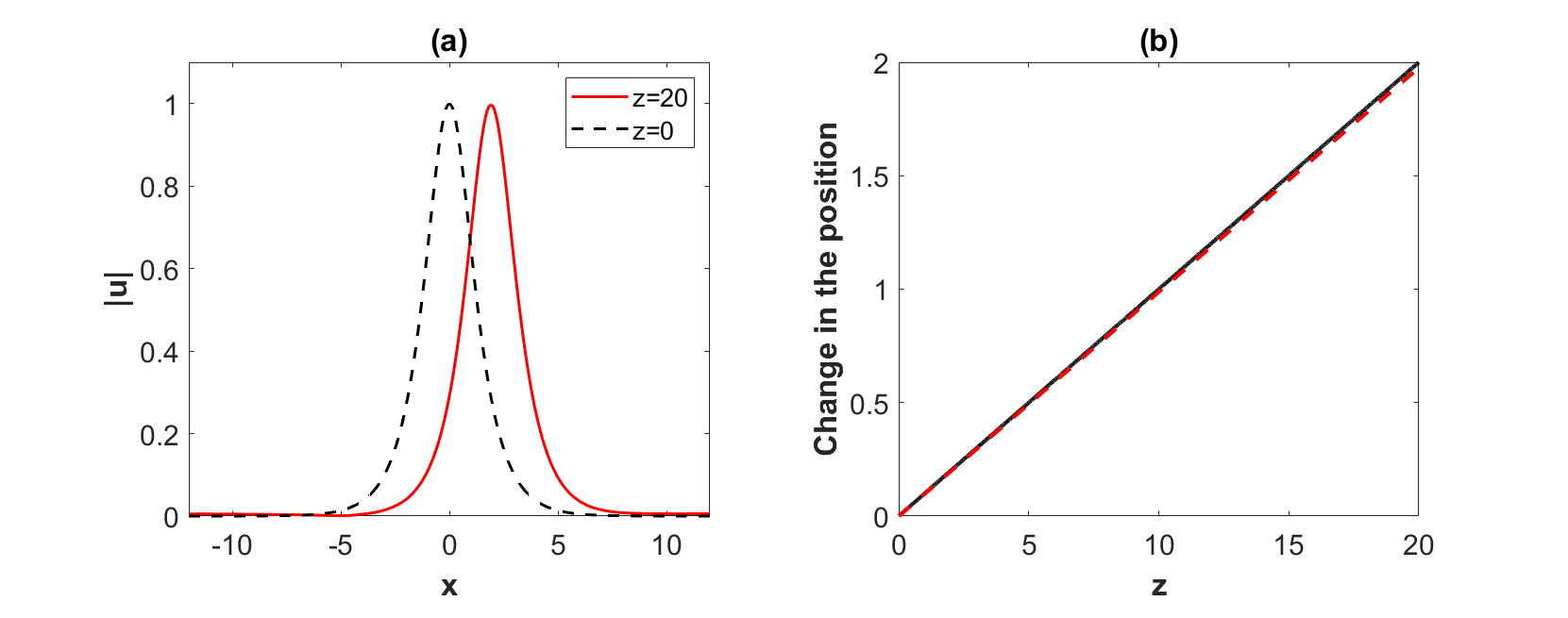}
\centering
\caption{Evolution of the $\mathrm{sech}(x)$ soliton  under the self-steepening effect (with $s=0.1$). (a) The profile of the soliton at $z=0$ and $z=20$. (b) The amount of position shift (vs. distance) computed by perturbation analysis (black solid line)  and the split-step method (red dashed line).}
\label{sech_posdif_2i}
\end{figure}

\begin{figure}[H]
\includegraphics[width=0.9\textwidth,keepaspectratio=true]{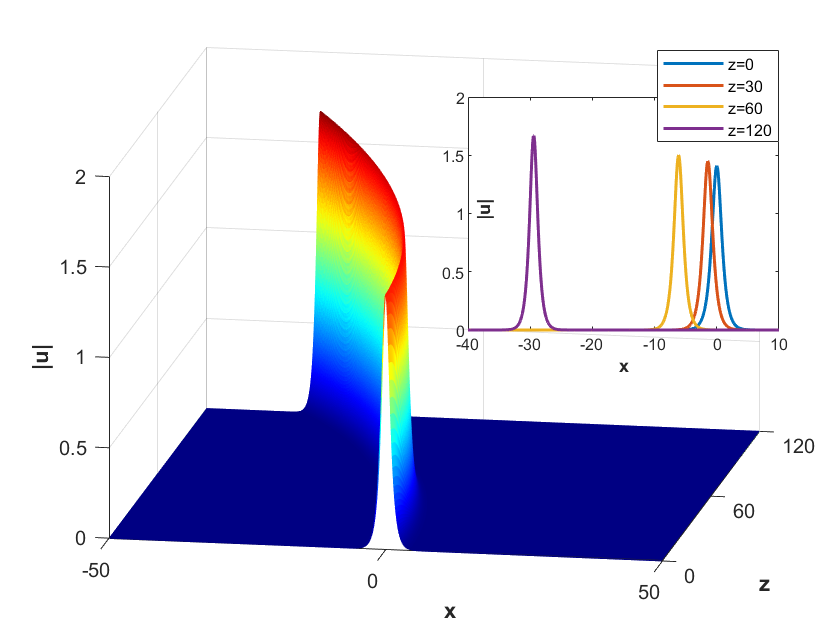}
\centering
\caption{Nonlinear evolution of the self-steepening soliton of Eq. (1) (with $\beta=1$ and $s=0.1$) in Eq. (1).}
\label{topview}
\end{figure}

\begin{figure}[ht]	\includegraphics[width=0.9\textwidth,keepaspectratio=true]{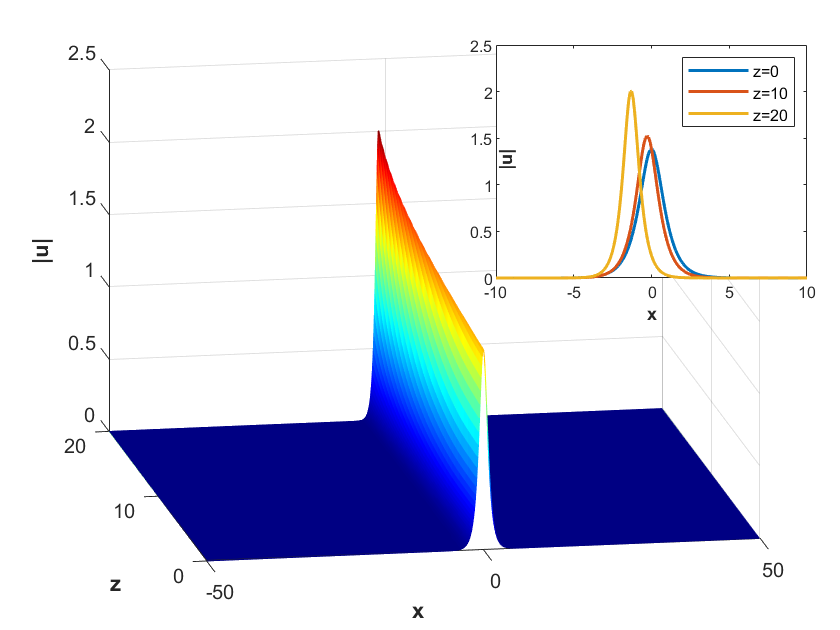}
	\centering
	\caption{Nonlinear evolution of the self-steepening soliton of Eq. (1) (with $\beta=1$ and $s=0.3$) in Eq. (1).}
	\label{fig:/ch5_evl_s_03}
\end{figure}

\begin{figure}[H]
\includegraphics[width=\textwidth,keepaspectratio=true]{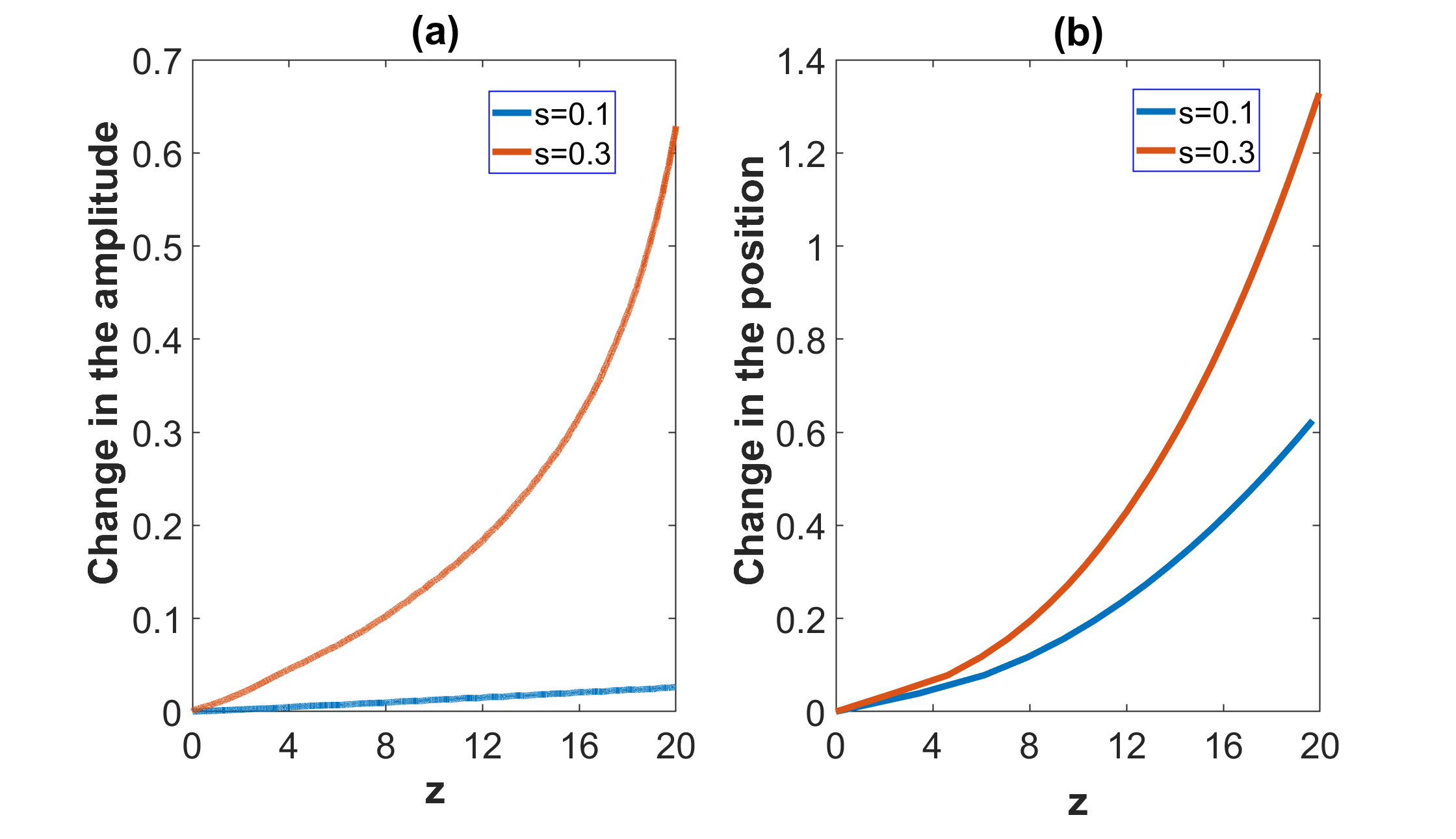}
\centering
\caption{The amount of change in peak amplitude and position of the self-steepening solitons (for $s=0.1$ and $s=0.3$) as a function of the propagation distance $z$.}
\label{change_pos_amp}
\end{figure}

It is well known that the NLS equation (MNLS equation with $s=0$) has stable fundamental soliton solutions.  However, including the self-steepening term in the NLS equation has some consequences on the stability of solitons. 
The MNLS equation admits solitons
\begin{equation}
u(z,x)=r(z) \operatorname{sech} r(z)[x-v(z)] e^{-i \delta(z)[x-v(z)]+i \sigma(z)},
\end{equation}
where
\begin{equation}
v=-\int_0^z \delta d s+\tau_0, \quad \sigma=\frac{1}{2} \int_0^z\left(r^2+\delta^2\right) d s+\sigma_0 .
\end{equation}
Here, $r$ denotes the soliton's amplitude, $v$ is the velocity, $\delta$  represents a frequency parameter, $\sigma_0$ is the soliton's initial phase, and $\tau_0$  indicates the amount of the soliton's position shift.
In \cite{yangbook}, the effect of self-steepening on an NLS soliton was analyzed by the perturbation
theory and  it's found that
\begin{equation}
\frac{d r}{d z}=\frac{d \delta}{d z}=\frac{d \sigma_0}{d z}=0, \quad \frac{d \tau_0}{d z}=s r^2.
\end{equation}
According to these results,
the main effect of self-steepening on the soliton is a position shift. This position shift increases linearly with distance in the amount of $\tau_0=sr^2z$.
To compare this phenomenon with the results of the split-step method, we numerically examine the evolution of an NLS soliton under the self-steepening effect.  Fig. \ref{sech_posdif_2i}(a) depicts the numerical evolution of the $\mathrm{sech}(x)$ soliton in Eq. (1) (with $s=0.1$). Fig. \ref{sech_posdif_2i}(b) then shows the amount of position shift (vs. distance) computed in two different ways, by perturbation analysis and the split-step method. The amount of position shift calculated by both methods shows good agreement, with very similar results.
In the rest of this paper, we examine the evolution simulations of self-steepening solitons obtained from Eqs. (1) and (2), not the effects of self-steepening on NLS solitons.
As part of our investigation into the nonlinear stability of self-steepening solitons, we obtain the soliton solution of Eq. (1) for $s=0.1$, and depict the nonlinear evolution of this soliton in Fig. \ref{topview}.  When the soliton moves from $z=0$ to $z=120$, the self-steepening effect shifts the position of the peak of the soliton from $x=0$ to $x=-29.39$ and increases the peak amplitude of the soliton from $1.41$ to $1.66$. Since the peak amplitude and position of the self-steepening soliton change it is considered to be nonlinearly unstable.
To gain a deeper understanding of the dynamic properties of self-steepening solitons, we find the soliton solution of Eq. (1) for $s = 0.3$, and depict the nonlinear evolution of the soliton in Fig. \ref{fig:/ch5_evl_s_03}. Although the propagation distance is reduced to $20$, the increase in the soliton's peak amplitude is significantly greater than in the case of $s=0.1$. In addition, there is a remarkable shift in the position of the soliton. 
Fig. \ref{change_pos_amp} shows the amount of amplitude increase and position shift of the peaks of the self-steepening solitons obtained for $s=0.1$ and $s=0.3$ during their propagation towards $z=20$.
It’s deduced from the figure that increasing the self-steepening coefficient increases the rate of change of the amplitude increase and the position shift.
\subsection{Self-steepening optical solitons in the real periodic potential}

\begin{figure}[ht]
\includegraphics[width=\textwidth,keepaspectratio=true]{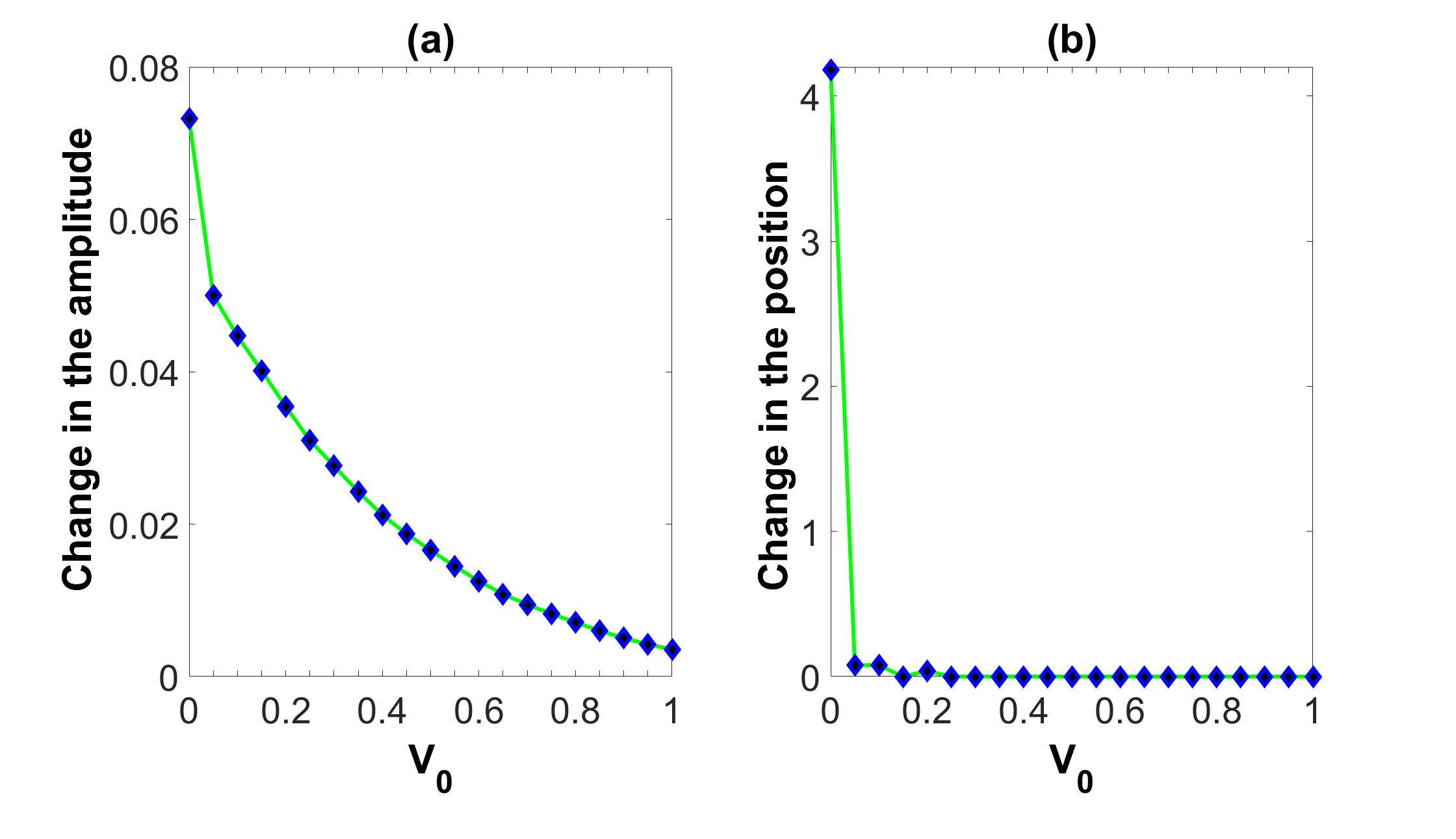}
\caption{The amount of change in peak amplitude and position of the self-steepening solitons (for $s=0.1$) between $z=0$ and $z=50$  as a function of the potential depth $V_0$.}
\label{ch5_evl_V0_3}
\end{figure}

\begin{figure}[ht]
\includegraphics[width=\textwidth,keepaspectratio=true]{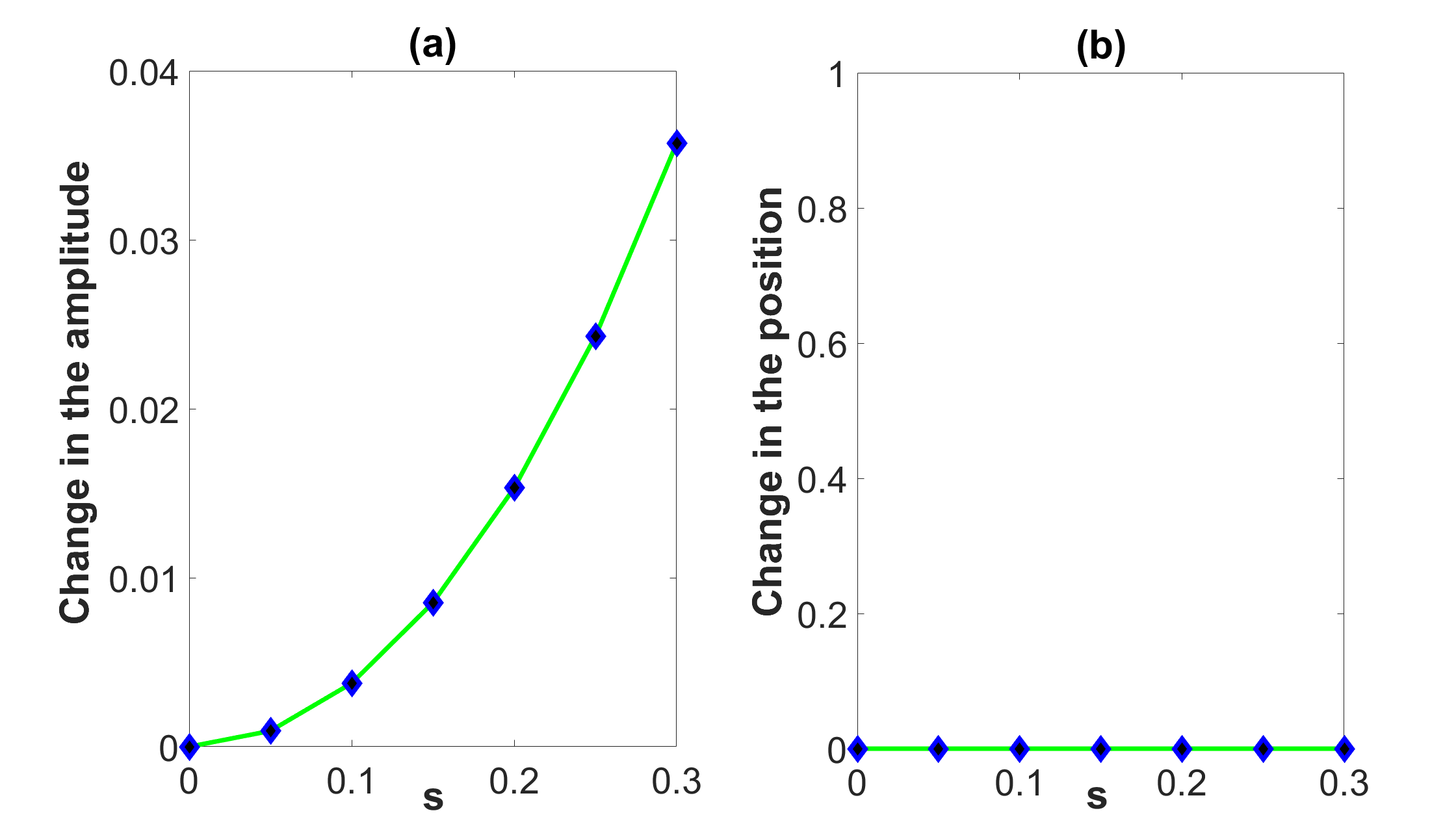}
\caption{The amount of change in peak amplitude and position of the self-steepening solitons between $z=0$ and $z=50$ as a function of the self-steepening coefficient $s$. The potential depth $V_0=0.7$.}
\label{ch5_evl_V0_4}
\end{figure}

 In this section, we aim to suppress the instability of self-steepening solitons by including the real component of the periodic $\mathcal{P}\mathcal{T}$-symmetric potential (3) in the MNLS equation. The effectiveness of this approach is investigated in Figs. \ref{ch5_evl_V0_3} and \ref{ch5_evl_V0_4}, where we analyze the nonlinear stability of self-steepening solitons in this real periodic potential.
In Fig. \ref{ch5_evl_V0_3},  the self-steepening coefficient is taken as $s=0.1$, and soliton solutions are found for various values of $V_0$ between $0$ and $1$. Then, these solitons are advanced up to    $z =50$ by the split-step method, and the amount of change in peak amplitude and position of these solitons between $z = 0$ and $z = 50$ is depicted. It is seen that as $V_0$ increases, the amount of amplitude increase of the solitons decreases. 
Furthermore, the change in the position of the solitons has completely disappeared for almost all $V_0>0$. In Fig. \ref{ch5_evl_V0_4}, the value of $V_0$ is fixed to $0.7$, and the amount of change in peak amplitude and position of self-steepening solitons are plotted as a function of $s$. While no position shift is observed even at a large value of $s$, such as $s=0.3$, the rate of increase in the amplitude of the solitons increases with increasing $s$ values.

\begin{figure}[H]
\includegraphics[width=\textwidth,keepaspectratio=true]{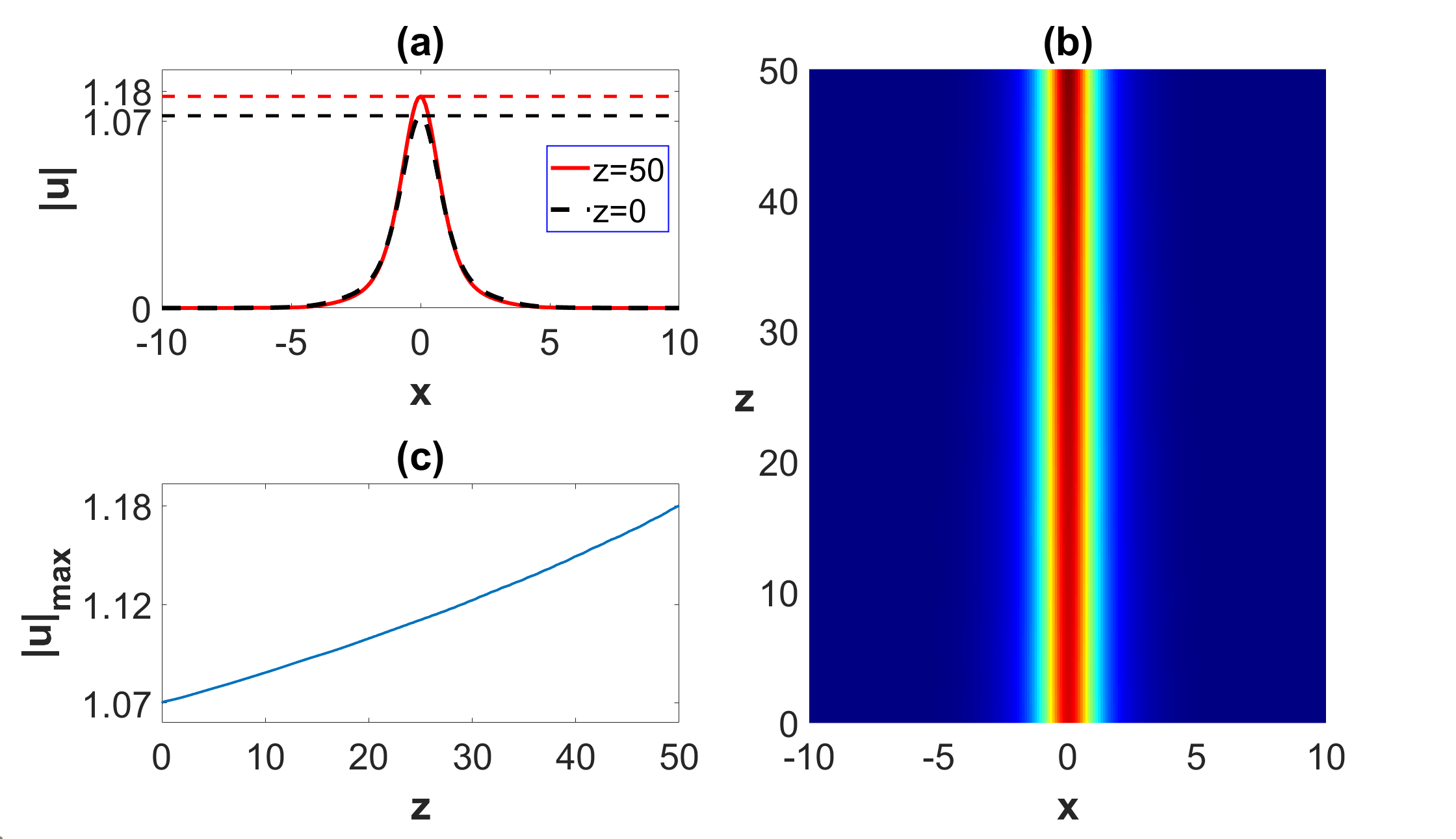}
\centering
\caption{Nonlinear evolution of the self-steepening soliton of Eq. (2) (with $\beta=1$, $s=0.3$, $V_0=0.7$, and $W_0=0$) in Eq. (2). (a) The profile of the soliton at $z=0$ and $z=50$. (b) Top view of the soliton propagation. (c) Peak amplitude  of the  soliton as a function of the propagation distance $z$.}
\label{ch5_evl_s_03_V0_07_2}
\end{figure}

 In Figure \ref{ch5_evl_s_03_V0_07_2}, we illustrate the nonlinear evolution of the self-steepening soliton for $s=0.3$ in the real periodic potential with the depth of the potential $V_0 = 0.7$. This figure shows that the position shift is suppressed by including the real periodic potential in the MNLS equation. On the other hand, despite a substantial reduction in the amplitude increase compared to the potential-free case, the soliton still exhibits a notable amplitude change. 

\subsection{Self-steepening optical solitons  in the periodic $\mathcal{P}\mathcal{T}$-symmetric  potential}

In the previous section, we have significantly suppressed the instability of self-steepening solitons by including the real part of the periodic $\mathcal{P}\mathcal{T}$-symmetric potential (3) in the MNLS equation.
Adding the real periodic potential to the MNLS equation eliminates the position shift and significantly reduces the amplitude increase. In this section, we investigate the use of the periodic $\mathcal{P}\mathcal{T}$-symmetric potential to further reduce the amplitude increase of self-steepening solitons during their propagation.

\begin{figure}[H]
\includegraphics[width=\textwidth,keepaspectratio=true]{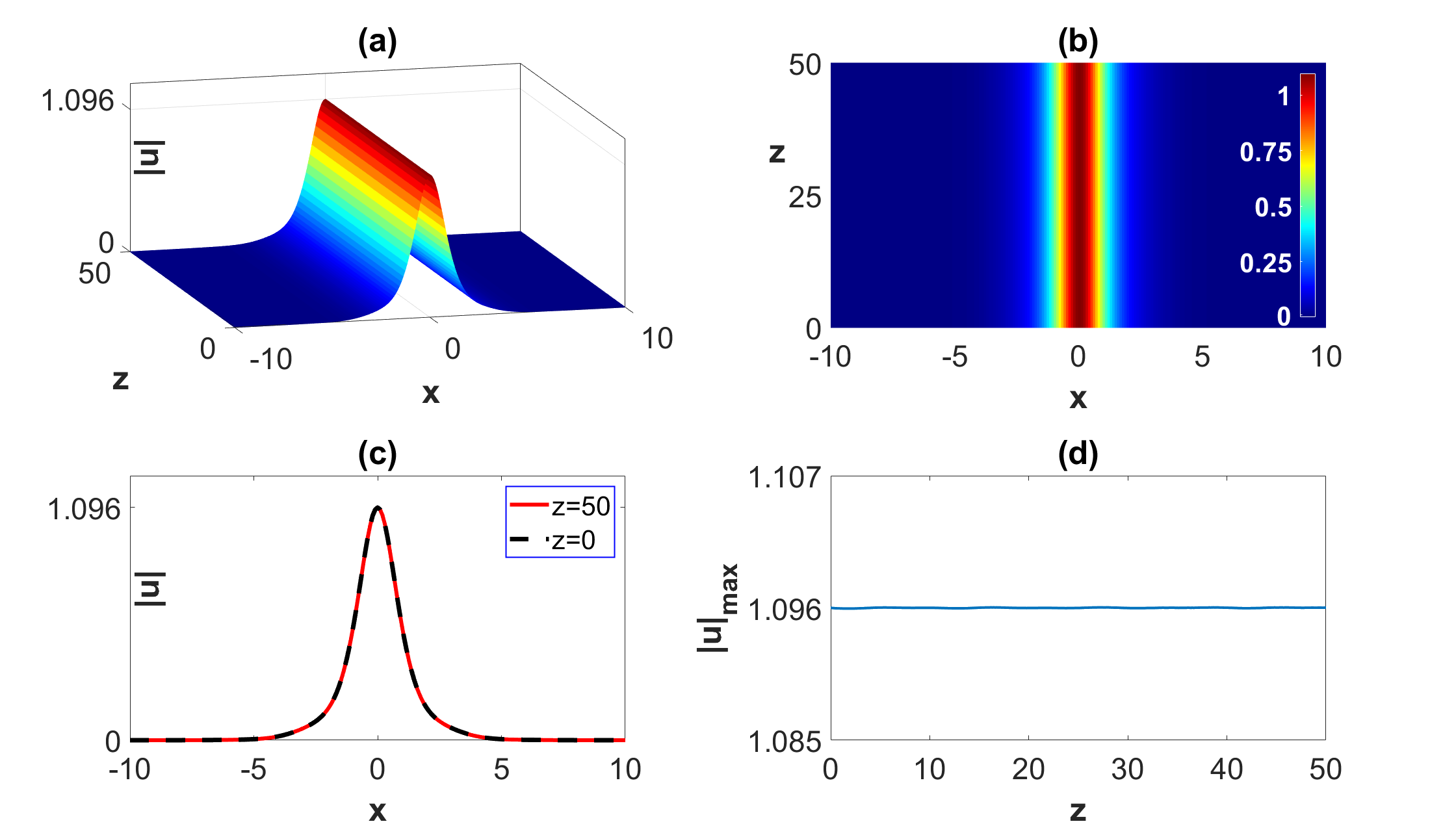}
\centering
\caption{Nonlinear evolution of the self-steepening soliton of  Eq. (2) (with $\beta=1$, $s=0.1$, $V_0=0.7$, $W_0=0.29135$) in Eq. (2). (a) Three-dimensional view. (b) Top view. (c) Optical pulses at $z=0$ and $z=50$. (d) Peak amplitude as a function of the propagation distance $z$.}
\label{ch5_evl_s_01_V0_07_W0_2}
\end{figure}

\begin{figure}[H]
\includegraphics[width=0.9\textwidth,keepaspectratio=true]{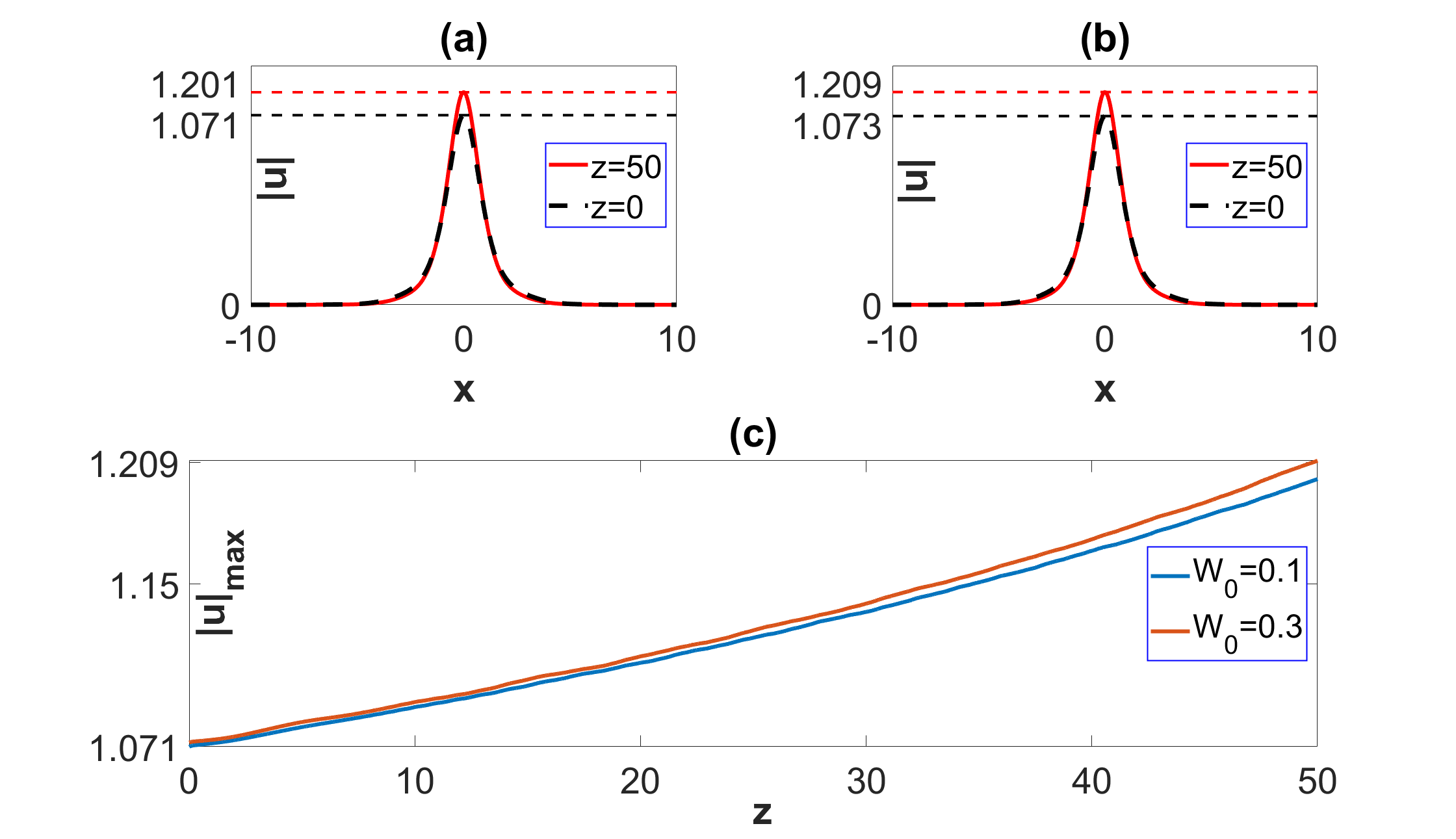}
\centering
\caption{Nonlinear evolution of the self-steepening soliton of  Eq. (2) (with $\beta=1$, $s=0.3$, $V_0=0.7$) in Eq. (2). (a) Optical pulses at $z=0$ and $z=50$ for $W_0=0.1$.
 (b) Optical pulses at $z=0$ and $z=50$  for $W_0=0.3$. (c) Amplitudes as a function of the propagation distance $z$.}
\label{Stab_AMPL_S03_V07_W4}
\end{figure}

\begin{figure}[ht]
\includegraphics[width=0.9\textwidth,keepaspectratio=true]{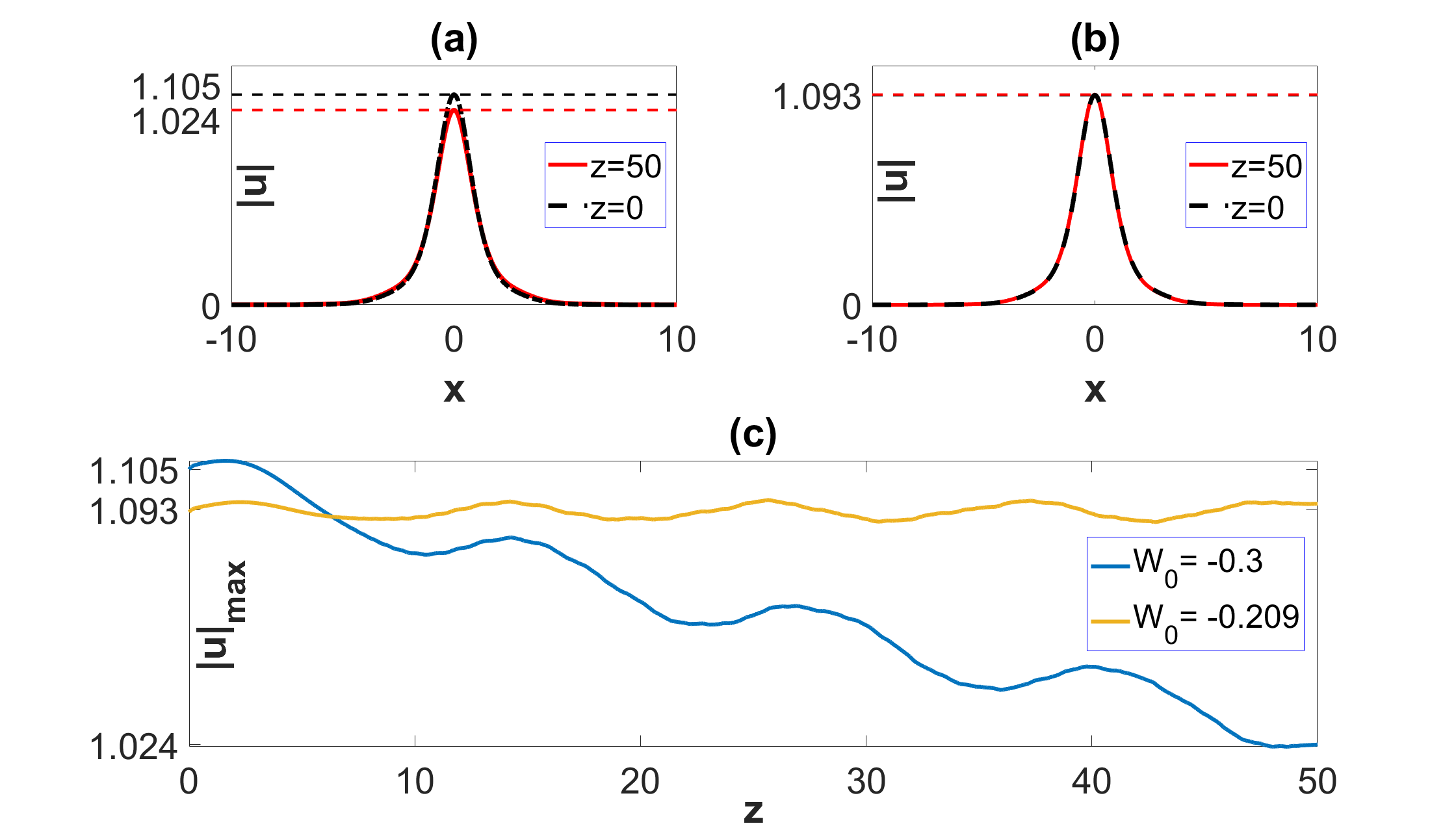}
\centering
\caption{Nonlinear evolution of the self-steepening soliton of  Eq. (2) (with $\beta=1$, $s=0.3$, $V_0=0.7$) in Eq. (2). (a) Optical pulses at $z=0$ and $z=50$  for $W_0=-0.3$.
 (b) Optical pulses at $z=0$ and $z=50$  for $W_0=-0.209$. (c) Amplitudes as a function of the propagation distance $z$.}
\label{Stab_AMPL_S03_V07_WN0_3}
\end{figure}

\begin{figure}[H]
\includegraphics[width=\textwidth,keepaspectratio=true]{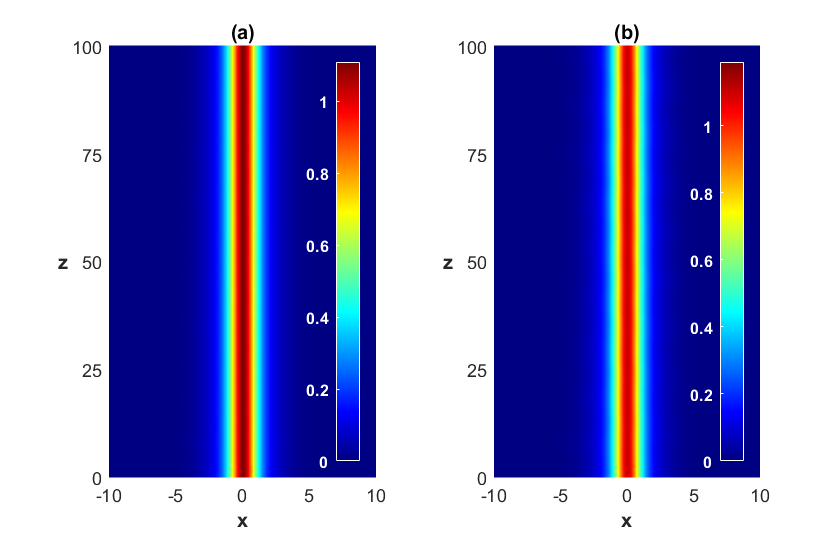}
\centering
\caption{(a) and (b) Nonlinear evolutions of the two solitons in Figs. 13(c) and 15(b) under $1\%$ random-noise  initial perturbations, respectively.}
\label{evol_perturbed}
\end{figure}

Fig. \ref{ch5_evl_s_01_V0_07_W0_2} shows the nonlinear
evolution of the self-steepening soliton of Eq. (2) with $s=0.1$, $V_0=0.7$ (depth of the real part of the potential), and $W_0=0.29135$ (depth of the imaginary part of the potential). The periodic $\mathcal{P}\mathcal{T}$-symmetric potential effectively eliminates the position shift phenomenon and reduces the amount of change in peak amplitude of the soliton between $z = 0$ and $z = 50$ to the order of $10^{-7}$.
Nevertheless, when the value of $s$  is increased to $0.3$, stability cannot be achieved for positive values of $W_0$. As an example, Fig. \ref{Stab_AMPL_S03_V07_W4} illustrates the soliton propagation for two different positive values of $W_0$. The increase in peak amplitude of optical pulses cannot be prevented when $s=0.3$ and $W_0 \geq 0$. In contrast, Fig. \ref{Stab_AMPL_S03_V07_WN0_3} shows the soliton propagation for two different negative values of $W_0$ when $s=0.3$.  As shown in the figure, a stable soliton can be achieved when the real and imaginary parts of the potential have depths of $V_0=0.7$ and $W_0=-0.209$, respectively. 
The two solitons in Figs. \ref{ch5_evl_s_01_V0_07_W0_2}(c) and \ref{Stab_AMPL_S03_V07_WN0_3}(b) are perturbed by $1\%$  random-noise perturbations, and their nonlinear
evolutions in Eq. (2) are displayed in Figs. \ref{evol_perturbed}(a) and \ref{evol_perturbed}(b), respectively. Figure \ref{evol_perturbed} shows that self-steepening solitons in Figs. \ref{ch5_evl_s_01_V0_07_W0_2}(c) and \ref{Stab_AMPL_S03_V07_WN0_3}(b) are stable against perturbations. 
Then, it can be concluded that stable self-steepening solitons can exist in the periodic $\mathcal{P}\mathcal{T}$-symmetric potential (\ref{pt}) even when the self-steepening coefficient $s$ has a large value of $0.3$.


\section{Conclusion}

In this study, we have conducted a numerical study on the existence, linear stability, and propagation dynamics (nonlinear stability) of self-steepening optical solitons in a periodic $\mathcal{P}\mathcal{T}$-symmetric potential (3). Firstly, we investigated the existence region of self-steepening solitons
of Eq. (2) according to the equation parameters.
Fig. 2(b) shows that the periodic $\mathcal{P}\mathcal{T}$-symmetric potential significantly extends the existence region of self-steepening solitons. Namely, self-steepening soliton solutions can be obtained for substantially larger values of
the self-steepening coefficient when the governing equation has the periodic $\mathcal{P}\mathcal{T}$-symmetric potential.
Secondly, we investigated the linear stability of self-steepening solitons of Eq (2)  by analyzing their linear stability spectra. Figs. 4 and 5 show that self-steepening solitons of Eq. (2) without the potential are linearly stable  even when the self-steepening coefficient ($s$) has a large value of $0.3$.  However, when the governing equation contains the periodic $\mathcal{P}\mathcal{T}$-symmetric potential, self-steepening solitons of Eq. (2) are considered weakly unstable. Because linear stability spectra of these solitons include some small positive real eigenvalues that are less than $10^{-3}$. Finally, we investigated the nonlinear evolution of self-steepening solitons. Figs. 7 and 8 show that self-steepening solitons of the MNLS equation (1) exhibit a position shift and an increase in peak amplitude during their evolution. By including the periodic $\mathcal{PT}$-symmetric potential (3) in the MNLS equation we achieved to suppress the nonlinear instability of self-steepening solitons.  It has been revealed that the real part of the periodic $\mathcal{P}\mathcal{T}$-symmetric potential eliminates the position shift and significantly reduces the amplitude increase (see Figs. 10-12). On the other hand, the imaginary part of the potential significantly contributes to the stability of self-steepening solitons by further reducing the amplitude increase (see Figs. 13 and 15).  

{In conclusion, we have demonstrated that the periodic $\mathcal{P}\mathcal{T}$-symmetric potential can stabilize the propagation of self-steepening optical solitons.
 This finding opens avenues for exploring how different $\mathcal{PT}$-symmetric potential configurations can manipulate soliton dynamics. Our ongoing research investigates these broader effects, including the confirmed stabilizing properties of certain configurations such as Wadati \cite{Yang_2014} and Scarf-II \cite{Chen_2020} potentials, as well as richer phenomena like symmetry breaking. Unveiling the diverse impacts of $\mathcal{PT}$-symmetric potentials on solitons experiencing higher-order effects, such as self-steepening, holds immense significance for future applications in controlling solitons' behavior across various fields.}

\appendix
\section{Fourier Collocation Method}
\label{appendix}
Firstly,   the infinite $x$-axis is truncated into a finite interval [$-L / 2, L / 2$], where $L$ is the length of the interval. Then  the eigenfunctions $[g, h]^T$ and functions $G_0$, $G_1$, $G_2$, $G_3$, $G_0^*$ ($G_4$), $G_1^*$ ($G_5$), $G_2^*$ ($G_6$), $G_3^*$ ($G_7$) in Eq. (15) are expanded into Fourier series:
\begin{equation}
\begin{gathered}
g(x)=\sum_n a_n e^{i n k_0 x}, \quad h(x)=\sum_n b_n e^{i n k_0 x}, \\
G_j=\sum_n c_n^{(j)} e^{i n k_0 x}, \quad j=0,1,...7,
\end{gathered}
\end{equation}
where $k_0=2 \pi / L$. Substituting these expansions into the eigenvalue problem (15) and equating the coefficients of the same Fourier modes, the following eigenvalue system for the coefficients $\left\{a_j, b_j\right\}$ will be obtained:
\begin{equation}
\begin{aligned}
&-\frac{1}{2}\left(k_0 j\right)^2 a_j+ \sum_n c_n^{(0)}  i(j-n) k_0  a_{j-n}+\sum_n c_n^{(1)} a_{j-n}\\
&+\sum_n c_n^{(2)}  i(j-n) k_0  b_{j-n}+\sum_n c_n^{(3)} b_{j-n}=-i \lambda a_j, 
\end{aligned}
\end{equation}
\begin{equation}
\begin{aligned}
& \frac{1}{2}\left(k_0 j\right)^2 b_j-\sum_n c_n^{(4)}  i(j-n) k_0  b_{j-n}-\sum_n c_n^{(5)} b_{j-n}\\
&-\sum_n c_n^{(6)}  i(j-n) k_0  a_{j-n}-\sum_n c_n^{(7)} a_{j-n}=-i \lambda b_j, 
\end{aligned}
\end{equation}
where $-\infty<j<\infty
$. By truncating the number of Fourier modes to $-N \leq j \leq N$, the infinite-dimensional eigenvalue problem  becomes the following finite-dimensional one:
\begin{equation}
i\left[\begin{array}{cc}
\frac{1}{2} D_2+C_0 D_1+C_1 & C_2 D_1+C_3 \\\\
-\left(C_6 D_1+C_7\right) & -\left(\frac{1}{2} D_2+C_4 D_1+C_5\right)
\end{array}\right]\left[\begin{array}{l}
A \\\\
B
\end{array}\right]=\lambda\left[\begin{array}{l}
A \\\\
B
\end{array}\right].
\end{equation}
Here
\begin{equation}
\begin{aligned}
{D}_1&=i k_0 \operatorname{diag}(-N,-N+1, \ldots, N-1, N), \\
{D}_2&=(i k_0)^2 \operatorname{diag}(-N,-N+1, \ldots, N-1, N)^2,\\\\
C_j&=\left(\begin{array}{ccccccc}
c_0^{(j)} & c_{-1}^{(j)} & \ldots & c_{-N}^{(j)} & & & \\
c_1^{(j)} & c_0^{(j)} & c_{-1}^{(j)} & \ddots & \ddots & & \\
\vdots & c_1^{(j)} & c_0^{(j)} & \ddots & \ddots & \ddots  & \\
c_N^{(j)} & \ddots & \ddots & \ddots & \ddots & \ddots & c_{-N}^{(j)} \\
& c_N^{(j)} & \ddots & \ddots & \ddots & \ddots & \vdots \\
& & \ddots & \ddots & \ddots & \ddots & c_{-1}^{(j)} \\
& & & & & & \\
& & & c_N^{(j)} & \ldots & c_1^{(j)} & c_0^{(j)}
\end{array}\right),\hspace{5px} j=0,1,2...7,\\\\
A&=\left(a_{-N}, a_{-N+1}, \ldots, a_N\right)^T, 
  \hspace{.8cm} B=\left(b_{-N}, b_{-N+1}, \ldots, b_N\right)^T.
\end{aligned}
\end{equation}
To solve the matrix eigenvalue problem (A.4), either the QR algorithm or the Arnoldi algorithm can be used.\\\\
\bibliographystyle{myvancouver}
\bibliography{article_revised}





\end{document}